\documentclass[10pt,conference]{IEEEtran}
\IEEEoverridecommandlockouts
\usepackage{cite}

\usepackage{amsmath,amssymb,amsfonts}
\usepackage{algorithmic}
\usepackage{graphicx,subfig}
\usepackage{textcomp}
\usepackage{xcolor}
\usepackage[letterpaper, top=0.75in, bottom=1.1in, left=0.63in, right=0.63in]{geometry}

\graphicspath{{./Figs/}}
\def\BibTeX{{\rm B\kern-.05em{\sc i\kern-.025em b}\kern-.08em
    T\kern-.1667em\lower.7ex\hbox{E}\kern-.125emX}}

\begin{document}

\title{Noise-Robust Radio Frequency Fingerprint Identification Using Denoise Diffusion Model\\

\thanks{J. Zhang is the corresponding author (email: Junqing.Zhang@liverpool.ac.uk). }
}

\author{
\IEEEauthorblockN{
Guolin~Yin\IEEEauthorrefmark{1},
Junqing~Zhang\IEEEauthorrefmark{1},
Yuan~Ding\IEEEauthorrefmark{2},
and Simon Cotton\IEEEauthorrefmark{3}
}

\IEEEauthorblockA{
\IEEEauthorrefmark{1}
Department of Electrical Engineering and Electronics, University of Liverpool, Liverpool, L69 3GJ, United Kingdom
}
\IEEEauthorblockA{
\IEEEauthorrefmark{2}
Institute of Sensors, Signals and Systems (ISSS), Heriot-Watt University, Edinburgh, EH14 4AS, United Kingdom
}
\IEEEauthorblockA{
\IEEEauthorrefmark{3}
Centre for Wireless Innovation (CWI), Queen’s University Belfast, Belfast, BT3 9DT, United Kingdom
}
}

\maketitle

\begin{abstract}
Securing Internet of Things (IoT) devices presents increasing challenges due to their limited computational and energy resources. Radio Frequency Fingerprint Identification (RFFI) emerges as a promising authentication technique to identify wireless devices through hardware impairments. 
RFFI performance under low signal-to-noise ratio (SNR) scenarios is significantly degraded because the minute hardware features can be easily swamped in noise.
In this paper, we leveraged the diffusion model to effectively restore the RFF under low SNR scenarios. Specifically, we trained a powerful noise predictor and tailored a noise removal algorithm to effectively reduce the noise level in the received signal and restore the device fingerprints. 
We used Wi-Fi as a case study and created a testbed involving 6 commercial off-the-shelf Wi-Fi dongles and a USRP N210 software-defined radio (SDR) platform. We conducted experimental evaluations on various SNR scenarios. The experimental results show that the proposed algorithm can improve the classification accuracy by up to 34.9\%.
\end{abstract}


\section{Introduction}
RFFI is an emerging authentication approach by using implicit hardware impairments in wireless transmitters, such as oscillators, mixers, and power amplifiers~\cite{zhang2021radio,wang2016wireless}. These impairments come from the manufacturing process which will deviate the nominal values of the hardware components slightly from their specifications.
The impairments are unique and can be extracted as device identifiers, in a similar manner to the biometric fingerprint authentication. As RFFI exploits the existing hardware impairments, this technique can be implemented solely at the receiver end but does not require any modification to the devices under test (DUTs). Therefore, it can be readily applied to any IoT networks.

Deep learning has been widely adopted for RFFI~\cite{shen2023deep,he2023channel} thanks to its superior feature extraction capability and classification capability. As indicated in~\cite{merchant2018deep, cekic2021wireless, yin2024multi}, the convolutional neural network (CNN) can significantly enhance the fingerprinting performance, outperforming the traditional method which uses hand-crafted features. Various deep learning techniques have been proposed to address challenges in RFFI. In \cite{kong2024towards}, the authors proposed a supervised contrastive learning-based method to address the channel variation. In ~\cite{shen2023toward}, the authors proposed a Transformer-based structure for overcoming the variable input length problem. The adversarial training method was proposed in~\cite{shen2023towards} to remove the impact of the receiver variations in RFFI.

SNR plays an essential role in RFFI. As the hardware fingerprints are subtle, they can be easily buried in noise. For example, the RFFI accuracy dropped from $97.1\%$ to $23.12\%$ in~\cite{kong2024towards} when SNR drops from 40 dB to 5 dB. Also, as reported in~\cite{he2024radio}, the RFFI performance dropped below $20\%$ when the SNR was around 20 dB.
The noise problem poses significant challenges for reliable RFFI. 
Data augmentation and collaborative identification~\cite{shen2023toward} can be employed to improve the noise robustness of RFFI systems. Data augmentation improves the diversity of the training datasets and collaborative identification integrates the deep learning predictions from multiple packets~\cite{shen2023toward} or receivers~\cite{shen2023towards}.
However, the RFFI accuracy under low SNR is still limited.

Diffusion models (DMs)~\cite{ho2020denoising, song2020denoising} have recently emerged as a powerful class of generative models, particularly effective for data generation tasks. They can generate new data by iteratively denoising Gaussian noise in a probabilistic manner. In addition, DMs can also be employed for denoising. By initializing the denoising process with a noisy input instead of pure Gaussian noise, the trained model can iteratively remove noise in a deterministic fashion. However, there is no research on using DMs for denoising in RFFI.

In this paper, we employed the DMs to train a powerful noise predictor for restoring the RFF. To adapt the pretrained noise predictor to the RFFI system, we proposed an SNR mapping algorithm that enables the noise predictor to accurately remove noise and restore RFF from noisy observations. 
We created a testbed consisting of commercial off-the-shelf Wi-Fi dongles as DUTs and a USRP N210 SDR platform as the receiver. Experimental evaluation was carried out. The technical contributions of this paper are summarized as follows:
\begin{itemize}
    \item We adapted the DM for denoising in the RFFI system to enable effective noise removal and restore RFF degraded by noise. To the best of the authors’ knowledge, this is the first work to apply DM for denoising within RFF.
    \item We proposed a Transformed-based RFFI approach by integrating the noise predictor. We designed an SNR mapping method, which allows the noise predictor to remove noise accurately and recover the underlying RFF in a computationally efficient manner.
    \item We experimentally evaluated the performance of the proposed system across a range of SNR levels. The experimental results demonstrated that the proposed system significantly enhanced identification accuracy, particularly under low SNR conditions. Specifically, the classification performance improved by 34.9\% at an SNR of 0 dB.
\end{itemize}

The rest of this paper is organised as follows. Section~\ref{sec:overview} presents the overview of the proposed RFFI system. Section~\ref{sec:diff} presents the technical details of the proposed diffusion model based denoising method. The details of the classifier used in this work are presented in Section~\ref{sec:classifier}. The experimental evaluation is presented in Section~\ref{sec:exp}. Finally, Section~\ref{sec:conclude} concludes this paper.

% \subsection{AWGN Channel}

% In this paper, we consider about the Additive White Gaussian Noise (AWGN) channel, which is a fundamental model used in wireless communication theory to represent the effect of random noise on a signal as it propagates through a medium. In this channel model, the transmitted signal is corrupted by the addition of a white noise process with a Gaussian distribution. Mathematically the received signal over the AWGN channel can be expressed as:
% \begin{equation}\label{eqn:awgn}
%     y = \mathcal{F}(x) + \epsilon,
% \end{equation}
% where $\mathbf{x}$ is the transmitted signal, $\mathcal{F}$ denotes the effect of the transmitter impairment and $\epsilon$ is the Gaussian noise. The presence of noise introduces uncertainty which makes it challenging to identify unique transmitter characteristics accurately. To effectively study the impact of noise on RFF it is essential to consider a controlled environment where other channel impairment, e.g., multipath fading, is minimized.  

\section{System Overview}\label{sec:overview}
\begin{figure}
    \centering
    \includegraphics[width=1\linewidth]{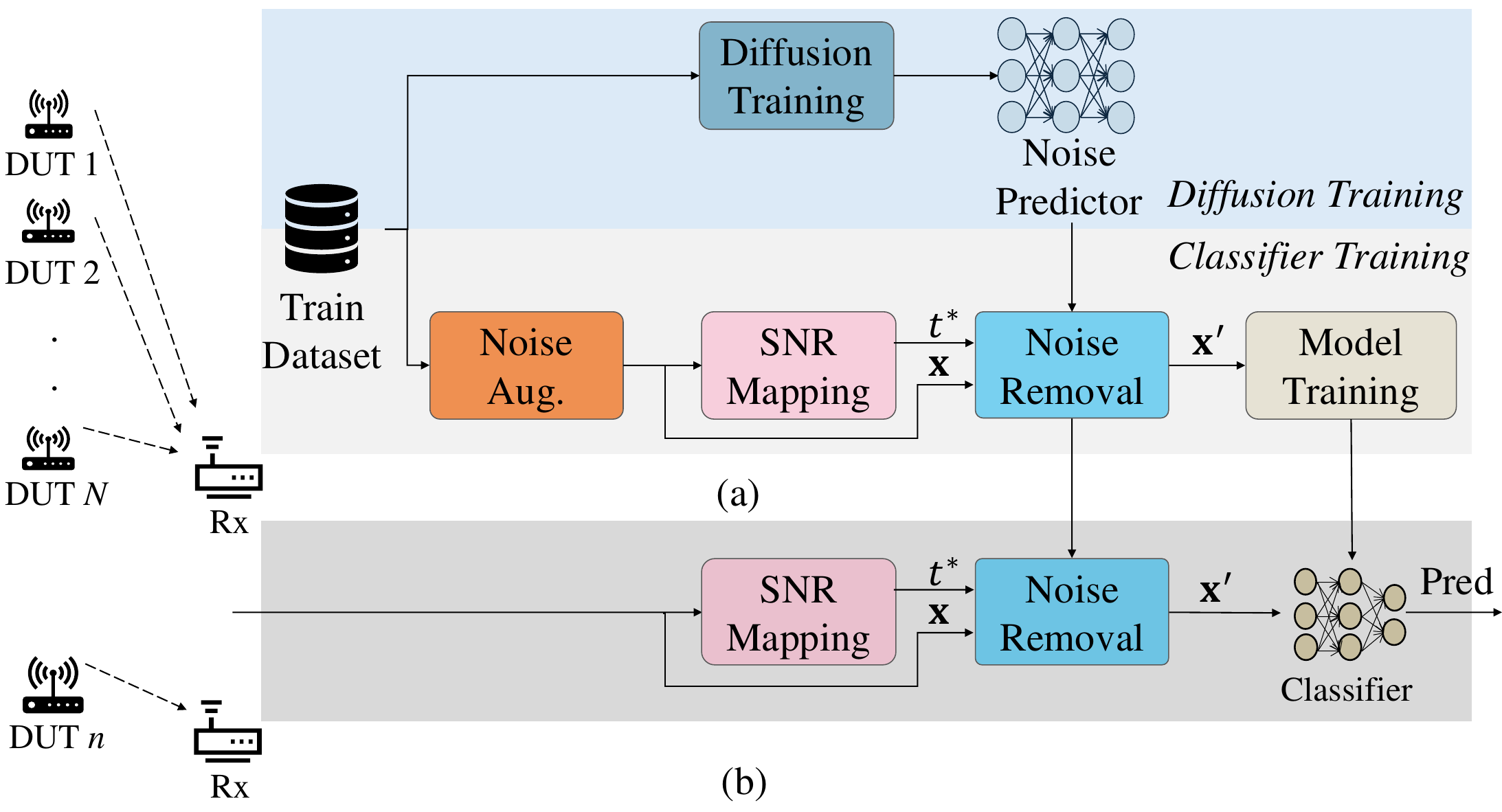}
    \caption{System overview. (a) Model training stage. (b) Inference stage.}
    \label{fig:systemoverview}
\end{figure}

Figure \ref{fig:systemoverview} depicts the system overview, which includes the model training and inference stages. During the training stage, each DUT sends packets that are captured by the receiver, forming a training dataset. This dataset is then utilized for DM training which yields a noise predictor model for noise removal. A classifier is also trained to identify the DUTs.

% \subsection{Preprocessing}
% For all two stages, after the signal is captured by the receiver, the signal preprocess scheme will be applied to the signal to prepare it for analysis and RFF extraction. The following preprocessing methods were applied in this work:

% \begin{itemize}
%     \item \textbf{Preamble extraction} isolates the preamble part of the signal, which in IEEE 802.11a consists of 320 sampling points over $16 \mu s$. 
%     \item \textbf{CFO compensation} enhances system stability by correcting frequency mismatches between the transmitter and receiver oscillators caused by temperature fluctuations and oscillator drift. According to previous studies [x], CFO is not a reliable feature for RFF.
%     \item \textbf{RMS normalization} scales the extracted signal to have a consistent root mean square (RMS) amplitude. This ensures that the deep learning model does account for the power fluctuation in RFF.
% \end{itemize}

\subsection{Model Training Stage}
\subsubsection{Diffusion Model Training} 
The objective is to develop a robust noise predictor capable of predicting noise contained in signal and then use it to recover the RFF features of transmitted signals distorted by noise. 
The DM training involves two steps, i.e., forward process and reverse process. The forward process involves adding noise to the training signals through a series of timesteps, whereas in the reverse process, a deep learning model is trained to learn how to remove the noise from the noisy observation and recover the original clean signals. The technical details will be presented in Section~\ref{sec:diff}.

\subsubsection{Classifier Training} The objective is to develop a model capable of extracting unique RFF features from the denoised signals produced by the DM trained in the previous step. 
%To enable the classifier to adjust to varying noise levels in signals denoised by the DM-trained noise predictor, we propose enhancing performance through two methods. 
First, we incorporate noise augmentation to increase the diversity of the training dataset. Second, we propose SNR mapping to determine the optimal denoising step $t^*$ based on the noise conditions.
Then, in the noise removal step, the noise predictor receives the noisy signal $\mathbf{x}$ along with the optimal timestep $t^*$. It processes these inputs to produce a denoised signal, which is then used to train a classifier to identify the RFF. The details of the classifier training will be presented in Section~\ref{sec:classifier}.

\subsection{Inference Stage}
In the inference stage, the system processes the incoming signals to identify the DUT. The receiver will first capture the packet that is sent from the DUT. The signal will be extracted along with the SNR value of the current signal. The SNR mapping module will estimate the optimal step for the subsequent noise removal procedure. Upon the received signal and SNR value, the noise predictor will remove noises and restore the essential RFF. 
The classifier will make a prediction on the denoised signal to predict the identity of the DUT.

\section{Restore RFF Feature with Diffusion Model}\label{sec:diff}
In this section, we first explain how to train the DM to obtain a noise predictor. Next, we detail the noise removal block, which uses the noise predictor to restore signals from noisy observations. 

\subsection{Diffusion Model Training}

\subsubsection{Forward Process}
The forward process~\cite{ho2020denoising} of the DM training aims to emulate the representation of a signal $\mathbf{x}$ in different levels of SNR by progressively adding Gaussian noise to the original data over a series of discrete timesteps.  

\begin{figure}[!t]
    \centering
    \includegraphics[width=0.9\linewidth, height=7.2cm]{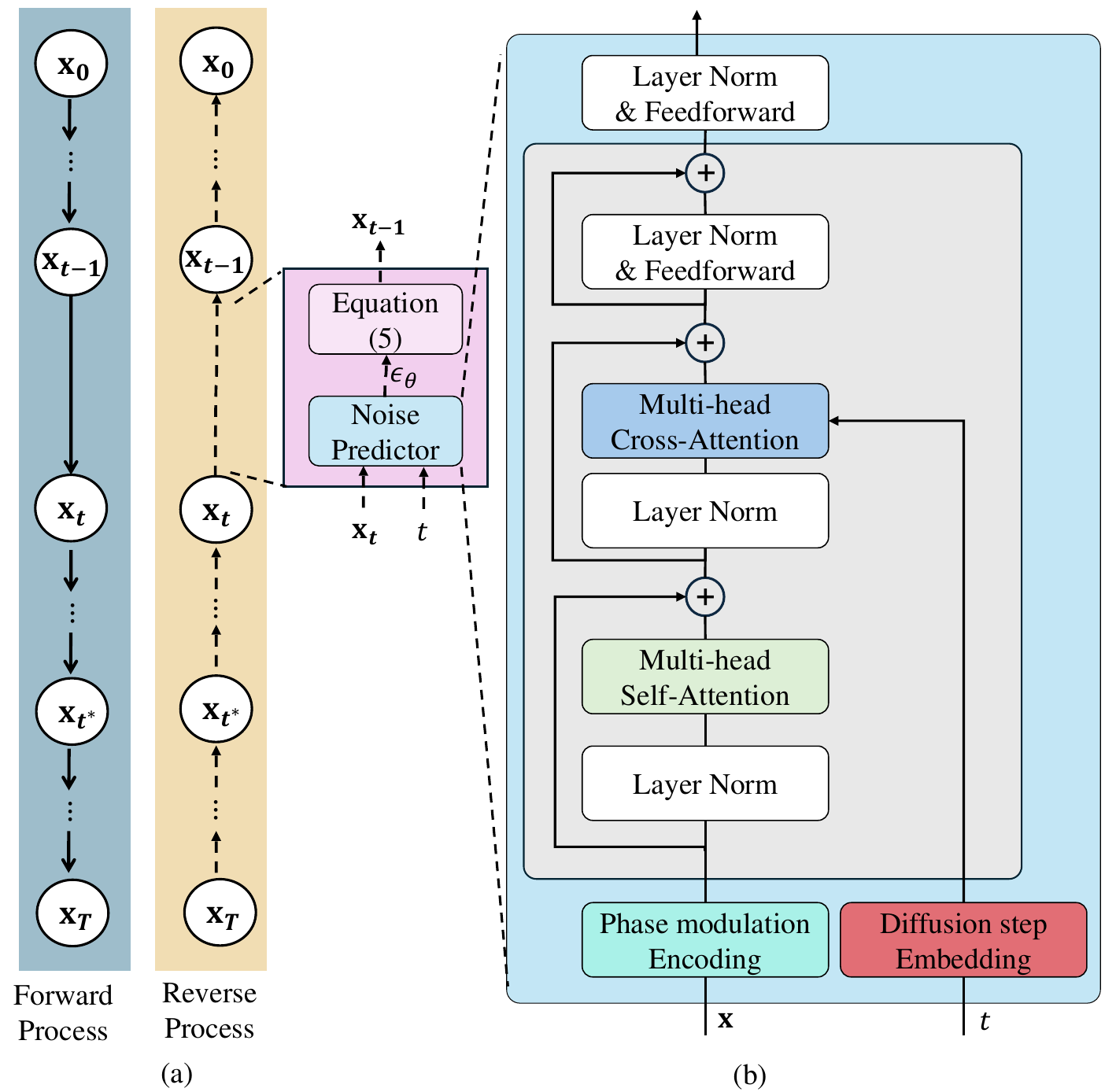}
    \caption{(a) The forward and reverse process. (b) The structure of noise predictor. }
    \label{fig:HDT}
\end{figure}

As shown in Fig.~\ref{fig:HDT}(a), the forward process will produce a series of noisy versions, $\left\{\mathbf{x}_t\right\}_{t=1}^T$, of the original input, $\mathbf{x_0}$. The noise power at each step is controlled by the variance schedule denoted as $\left\{\alpha_t \in(0,1)\right\}_{t=1}^T$ and $\alpha_t = 1 - \beta_t$, then the forward process is defined as:
\begin{equation}
q(\mathbf{x}_t|\mathbf{x}_{t-1}) = \mathcal{N}\left(\mathbf{x}_t; \sqrt{\alpha_t} \mathbf{x}_{t-1}, (1 - \alpha_t) I\right),
\end{equation}
where $\mathcal{N}$ denotes a Gaussian distribution with mean $\sqrt{\alpha_t} \mathbf{x}_{t-1}$ and variance $1 - \alpha_t$. Furthermore, the forward process can be expressed in a closed-form equation that directly computes the noisy signal $\mathbf{x}_t$ from the original signal $\mathbf{x}_0$, given as
% \begin{equation}
\begin{equation}
\label{eqn:forward}
\mathbf{x}_t=\sqrt{\bar{\alpha}_t} \mathbf{x}_{0}+\sqrt{1-\bar{\alpha}_t} \epsilon,
\end{equation}
where $\bar{\alpha}_t=\prod_{i=1}^t \alpha_i$ represents the cumulative product of noise scheduling parameters $\alpha_i$. The SNR of the signal at the timestep $t$, $\gamma_t$, can be calculated as 
\begin{equation}\label{eqn:snrmap}
    \gamma_t = \frac{\bar{\alpha}_t \, P_{\text{s}, 0}}{\bar{\alpha}_t \, P_{\text{n}, 0} + (1 - \bar{\alpha}_t)},
\end{equation}
where $P_{\text{s}, 0}$ is the signal power of $\mathbf{x}_0$ and $P_{\text{n}, 0}$ is the power of noise in the original signal $\mathbf{x}_0$. 

% Denote the timestep $t \in [0,T]$ as the index of the forward diffusion process which can be expressed mathematically:
% \begin{equation}\label{eqn:forward_step}
%     {x}_t=\sqrt{\alpha_t} {x}_{t-1}+\sqrt{1-\alpha_t} {\epsilon} 

% \end{equation}
% where $\alpha_t$ is the noise scheduling parameter which controls how much noise is added at each timestep $t$. By adjusting $\alpha_i$ we can adjust the signal-to-noise (SNR) ratio at each timestep $t$ during the forward process.

This process is crucial for subsequent reverse process to train robust noise predictor. By understanding how noise progressively distorts the signal, the model learns to effectively reverse this process, removing noise while preserving the essential RF features necessary for accurate device identification.

\subsubsection{Reverse Process}

The objective of the reverse process is to systematically remove the noise added during the forward diffusion process, thereby reconstructing the original clean signal from its noisy counterparts. In the reverse process, a deep neural network is employed to predict the noise added in the previous timestep. 

In this work, we modified the Hierarchical Diffusion Transformer (HDT) proposed in~\cite{chi2024rf}, originally tailored for RF signal generation, to serve as the backbone model for the noise predictor, $f_\theta$, parameterized by $\theta$. The model structure is shown in Fig.~\ref{fig:HDT}(a). Different from the original HDT, we removed class embedding as the class information during the denoising stage is not available in RFFI. 
The core block is a Transformer-based structure shown in the gray box in Fig.~\ref{fig:HDT}(a). This differs from the original Transformer encoder~\cite{vaswani2017attention}, as it introduced multi-head cross-attention that takes the embedded $t$ as input and allows the model to estimate the current noise level. The multi-head self-attention (MHSA) captures autocorrelation features from the noisy input and extracts high-level representations implicit in the signal. The gray box in 
the diffusion step embedding encodes the current diffusion step $t$ which will be used as input of MHSA.
The Phase Modulation Encoding takes a complex-valued signal as input and encodes positional information in the complex domain.

% Adaptive Layer Normalization (adaLN) replaces standard layer normalization by regressing scale and shift parameters from the diffusion step embedding t, effectively incorporating diffusion step information into the model's normalization layers.

As shown in the Fig~\ref{fig:HDT}(b). The reverse process involves learning a series of denoising steps that progressively refine the noisy data $\mathbf{x}_T$ to the original clean signal $\mathbf{x}_0$. Mathematically, the reverse process is modeled as a sequence of conditional probability distributions $p_\theta(\mathbf{x}_{t-1}|\mathbf{x}_t, t)$, where each distribution aims to predict the noise $\epsilon$ added from the previous noisy signal $\mathbf{x}_{t-1}$ during the forward process. 
The noise predicted by the noise predictor at step $t$ is $\epsilon_\theta\left(\mathbf{x}_t, t\right)$, which is denoted as $\epsilon_\theta$ too for simplicity.
During the DM training, the parameter of noise predictor, i.e., $\theta$, is updated by minimizing the mean squared error between the actual noise $\epsilon$ and the predicted noise $\epsilon_\theta$. The loss function is given as
\begin{equation}
    \mathcal{L}(\theta)=\mathbb{E}_{t, \mathbf{x}_0, \epsilon}\left[\left\|\epsilon-\epsilon_\theta\left(\mathbf{x}_t, t\right)\right\|^2\right].
\end{equation}

The previous signal after the denoising can be computed as
\begin{equation}
    \mathbf{x}_{t-1}=\sqrt{\alpha_{t-1}} \left(\frac{\mathbf{x}_{t}-\sqrt{1-\bar{\alpha}_{t}} \epsilon_{\boldsymbol{\theta}}}{\sqrt{\bar{\alpha}_{t}}}\right)+\sqrt{1-\bar{\alpha}_{t-1}}
    \boldsymbol{\epsilon}_{\boldsymbol{\theta}}.
		\label{eqn:forward_snr}
\end{equation}

\subsection{Noise Removal}
The noise predictor obtained during the DM training phase is used to remove noise from the received noisy signal. The generative task starts by removing the noise from pure Gaussian noise. Differently, in our task, we start from a noisy signal which is from intermediate, denoted as state ``$x_{t^*}$", as shown in the Fig.~\ref{fig:HDT}(a). Denoted optimal intermediate step as $t^*$, which will be detailed in Section~\ref{sec:snrmap}.
In this phase, the received signal is refined iteratively over $t^\prime$ steps.
In the DDPM framework, the denoising process, also referred to as the sampling process, is performed iteratively from step $t^*$ to step $t_0$, encompassing the total steps $t^*$, i.e., $t^\prime = t^*$. This method proves to be both time-consuming and computationally intensive, posing challenges for real-time processing required in RFFI systems. 

To achieve efficient and precise noise removal, we adopted the step-skipping method, which is originally proposed in Denoising Diffusion Implicit Models (DDIM)~\cite{song2020denoising}. In this approach, the number of timesteps is selected such that $t^\prime < t^*$. The timestep $t$ is determined as follows:

\begin{equation}
    t_i = t^* - i\cdot\Delta t
\end{equation}
where $\Delta t = \frac{t^*}{t^\prime}$ and $i = 1,2,...,t^\prime$. 

DDPM~\cite{ho2020denoising} employs a probabilistic sampling method by adding noise at each denoising step. In contrast, DDIM sampling~\cite{song2020denoising} used in our approach focuses on deterministically recovering the original signal without introducing additional noise. This distinction is crucial for signal recovery tasks, where the aim is to accurately reconstruct the original signal rather than to generate new variations. During the noise removal phase, the $f_\theta$ obtained in minimise~(\ref{eqn:forward_snr}) is used to predict the noise $\epsilon_\theta$ based on the current noisy observation $\mathbf{x}_t$ and timestep $t$. 

This deterministic denoising method ensures that noise is removed without additional randomness, which is important for refining the signal while preserving its RFF features necessary for accurate device identification. For simplicity, we denote the output of the noise removal as $\mathbf{x^\prime}$.

\section{Classifier Design}\label{sec:classifier}

%In this section, we present the key methods, noise augmentation and SNR mapping, that enable the classifier to train on the denoised signal of the noise predictor model. Additionally, we describe the architecture of the classifier.

\subsection{Noise Augmentation}
To enhance the robustness of the classifier against varying noise conditions, we employ the augmentation method proposed in~\cite{shen2023toward}. This approach involves the addition of Gaussian noise to different training batches to increase the diversity of the entire dataset. 
Denoting the initial SNR of a training batch as $\gamma_{init}$, the addition of Gaussian noise will adjust the batch SNR to a target value $\gamma_{tgt}$, where $\gamma_{tgt} < \gamma_{init}$. 
By increasing the diversity of the training dataset, the noise robustness can be improved.

\subsection{SNR Mapping}\label{sec:snrmap}
The SNR mapping step aims to determine the optimal timestep $t^*$ that corresponds to the current SNR of the received signal. This mapping ensures that the denoising operations are accurately aligned with the specific noise level present in the signal, allowing the model to apply the appropriate denoising step tailored to that noise intensity. Without SNR mapping, the noise predictor may insufficiently reduce the noise or overestimate the noise, obscuring the essential recovery of the RFF feature, which is necessary for accurate device identification.

To accurately determine the optimal $t^*$, we define the SNR map as $\gamma_{\text{map}} = \{ \gamma_0, \gamma_1, ..., \gamma_T \}$ that can be computed using~(\ref{eqn:snrmap}). We determine $t^*$ by finding the timestep in which the SNR value of the predefined noise schedule matches the current SNR, given as
\begin{equation}\label{eqn:optt}
    t^*=\arg \min _t|\gamma_{\text {map }}(t)-\gamma|,
\end{equation}
where $\gamma$ is the SNR of a signal input to the noise predictor.

\subsection{Classifier Model Structure}
\begin{figure}
    \centering
    \includegraphics[width=1\linewidth]{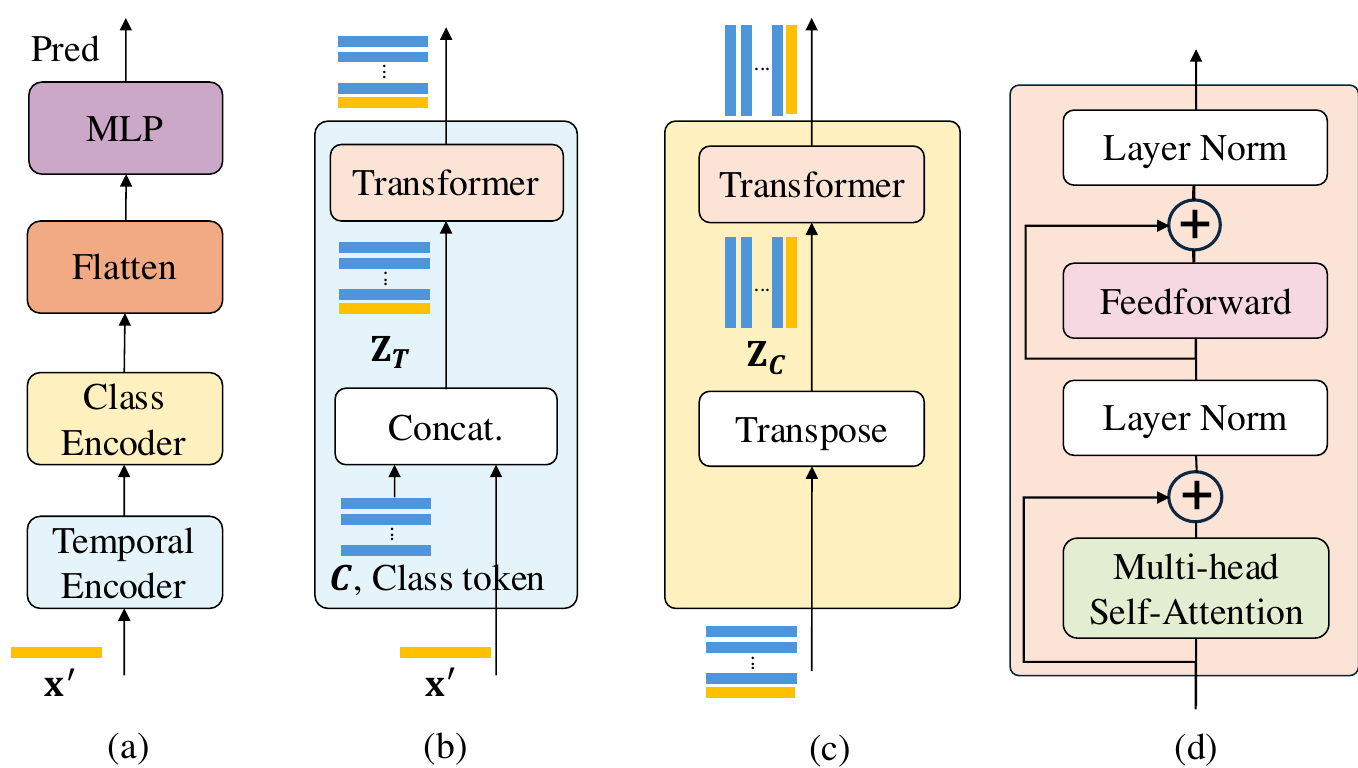}
    \caption{(a) The proposed Transformer-based classifier structure. (b) The temporal encoder. (c) The class encoder. (d) Backbone transformer encoder.}
    \label{fig:classifier}
\end{figure}

We employed a Transformer-based classifier as shown in Fig.~\ref{fig:classifier}(a), the classifier consists of the temporal encoder and the class encoder. For both the temporal and class encoders, the Transformer is used as the backbone.

\subsubsection{Temporal Encoder}
The temporal encoder processes the signal throughout the time dimension, capturing the inherent temporal patterns and dynamics. Inspired by the Vision Transformer (ViT)~\cite{dosovitskiy2020image}, a learnable class token is added to the input sequence of image patches to provide a global representation, aggregating information through self-attention for classification decisions. Expanding on this, we introduce multiple learnable class tokens, one for each class, as shown in Fig.~\ref{fig:classifier}(b). Class tokens enhance a model's capacity to aggregate temporal characteristics, enabling concurrent recognition of patterns unique to specific classes as well as temporal linkages within the signal. The concatenation is given as

% These class tokens are designed to learn and aggregate temporal features across the temporal dimensions. Each token focuses on capturing patterns and relationships relevant to its specific class, enabling the model to learn temporal dependencies for each class within the signal. 

% In the ViT, a learnable class token is concatenated to the input sequence of image patches to serve as a global representation that aggregates information from all patches through self-attention, ultimately allowing the model to make classification decisions based on the whole image context. 
% We adopted a similar concept of ViT, but instead of using a single class token, we used multiple class-specific tokens (one per class), as shown in Fig.~\ref{fig:classifier}(b), that can learn to aggregate temporal features across the input dimensions, enabling the model to simultaneously capture class-specific patterns and temporal relationships along the signal. Mathematically given as:
\begin{equation}
    \mathbf{Z}_T = concat(\mathbf{C}, \mathbf{x}^\prime ) \in \mathbb{R}^{(N+1) \times M},
\end{equation}
where $N$ denotes the total number of classes, $\mathbf{C} \in \mathbb{R}^{N \times M}$ represents the class tokens, $\mathbf{x}^\prime \in \mathbb{R}^{1 \times M}$ corresponds to the input vector for the classifier, and $M$ denotes the length of the signal.

\subsubsection{Class Encoder}
Following temporal feature extraction, the class encoder extracts relationships between different classes. 
The class encoder is designed to refine and aggregate the temporal features extracted by the temporal encoder to facilitate accurate classification. As shown in Fig.~\ref{fig:classifier}(c), the class encoder employs the Transformer block with the same structure as the temporal encoder to model the inter-class relationships and the extracted temporal features. Specifically, the class encoder takes the output of the temporal encoder as input and then performs a transpose on the input to obtain $\mathbf{Z}_C \in \mathbb{R}^{M \times (N+1)}$ to enable the Transformer to effectively capture class-wise dependencies.

% Specifically, the temporal features $\mathbf{Z}_T$ are transposed:
% \begin{equation}
%     \mathbf{Z}_C = transpose(\mathbf{Z}_T) \in \mathbb{R}^{M \times (N+1) },
% \end{equation}
% where $\mathbf{Z}_C  \in \mathbb{R}^{M \times (N+1) }$ is the input of the class encoder. 
The transpose operation will enable the Transformer to work over different class tokens to extract key features to distinguish different classes. This hierarchical processing, from temporal to class-specific encoding, enhances the model's ability to capture both temporal dependencies and inter-class relationships.

\subsubsection{Classification}

After the class encoder, the output feature map is passed through a flatten layer, which transforms the multidimensional tensor into a one-dimensional vector. This flattened vector serves as the input to the subsequent Multilayer Perceptron (MLP) responsible for classification. The classifier is trained using the cross-entropy loss.

\section{Experimental Evaluation}\label{sec:exp}

\subsection{Data Collection and Experiments Setup}
\begin{figure}[!t]
\centering
\subfloat[]{\includegraphics[width=0.6\linewidth, height=3cm]{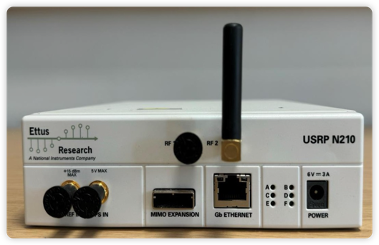}}
\subfloat[]{\includegraphics[width=0.2\linewidth, height=3cm]{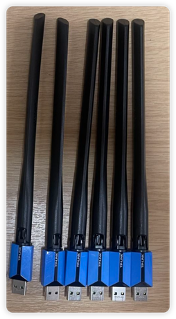}}
\centering
\caption{The experiment devices. (a) USRP N210. (b) TP-Link USB dongle.}
\label{fig:eqp}
\end{figure}
\subsubsection{Experiment Devices}
In this work, as shown in Fig.~\ref{fig:eqp}, we employed a USRP N210 SDR as the receiver and 6 Wi-Fi dongles as DUTs. During the data collection stage, the DUTs were plugged into a Linux laptop which was connected to a Wi-Fi router to create a packet stream. At the receiver end, the N210 was connected to a Linux laptop installed with Picosence~\cite{jiang2021eliminating} to obtain the baseband signal. The sampling rate was 20 Msamples/s.

\subsubsection{Datasets} 
In this work, we use IEEE 802.11 as a case study, specifically leveraging its legacy preamble which consists of Short Training Sequence (STS) and Long Training Sequence (LTS) for RFF extraction, as these standardized training sequences provide consistent and comparable features across all compliant devices. 

During the data collection, the USRP N210 and the WiFi dongles were placed about 1 meter apart with line-of-sight (LOS) present. Therefore, the SNR values of the collected packets all exceed 40~dB. 
The channel was kept stationary to maintain a fixed multipath channel.
Under this configuration, we can focus exclusively on noise effects. The different noise levels are emulated by noise augmentation by adding artificial noise in a simulation manner.

We collected 30,000 WiFi packets at 2.4~GHz on channel 13 for each DUT as the training dataset. For testing, we obtained 2,000 packets for each DUT on the same channel. 

\subsubsection{Training Configuration}
We conducted deep learning training on a PC equipped with an NVIDIA GeForce RTX 4090 GPU using PyTorch. The neural network parameters were optimized with the Adamax optimizer, starting with an initial learning rate of 0.0001 and a batch size of 32. We choose the linear spaced $\beta_t$ range from $1 \times 10^{-5}$ to $1.5 \times 10^{-3} $. During training, we halved the learning rate if the validation loss failed to improve for 20 consecutive epochs. Training was stopped when the validation loss showed no improvement for 30 epochs.

\subsection{Denoise Performance}
\begin{figure}[!t]
    \centering
    \includegraphics[width=1\linewidth, height=4cm]{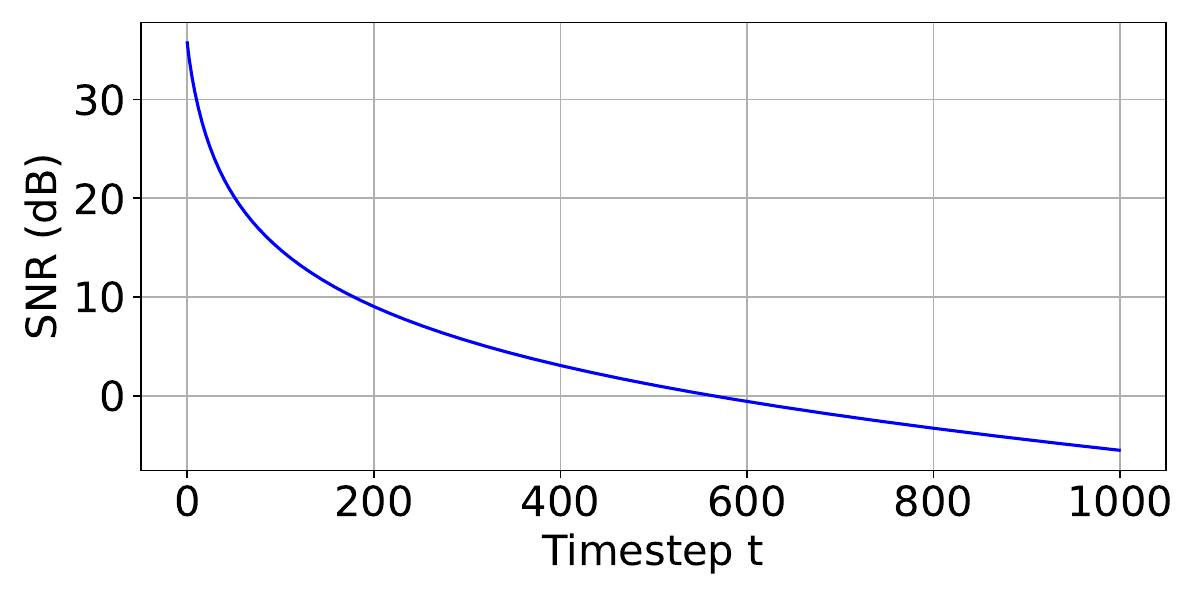}
    \caption{The noise schedule during diffusion model training.}
    \label{fig:snr_vs_t}
\end{figure}

The SNR schedule across timesteps $t$ can be computed using (\ref{eqn:snrmap}) and is illustrated in Fig.~\ref{fig:snr_vs_t}. This noise schedule ensures that the model is exposed to varying noise levels, allowing it to learn effective noise removal strategies across a wide range of signal quality conditions. The training process is designed to progressively denoise the received signals, preserving the essential features required for accurate RFF.

\begin{figure*}[!t]
\centering
\subfloat[]{\includegraphics[width=0.333\linewidth]{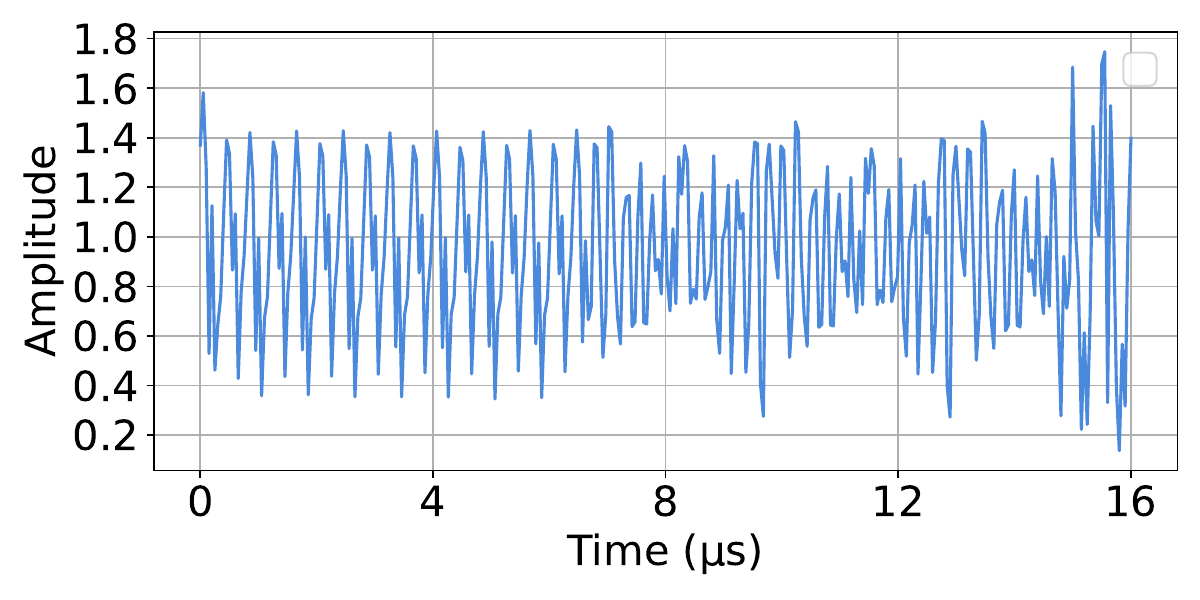}\label{fig:original}}
\subfloat[]{\includegraphics[width=0.333\linewidth]{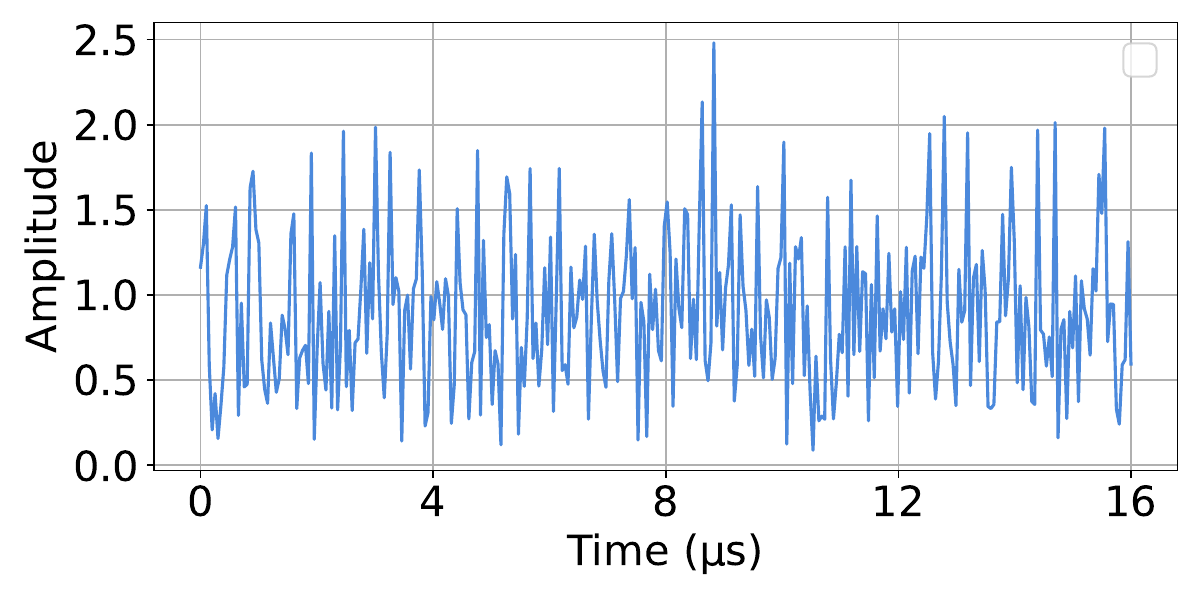}\label{fig:noised}}
\subfloat[]{\includegraphics[width=0.333\linewidth]{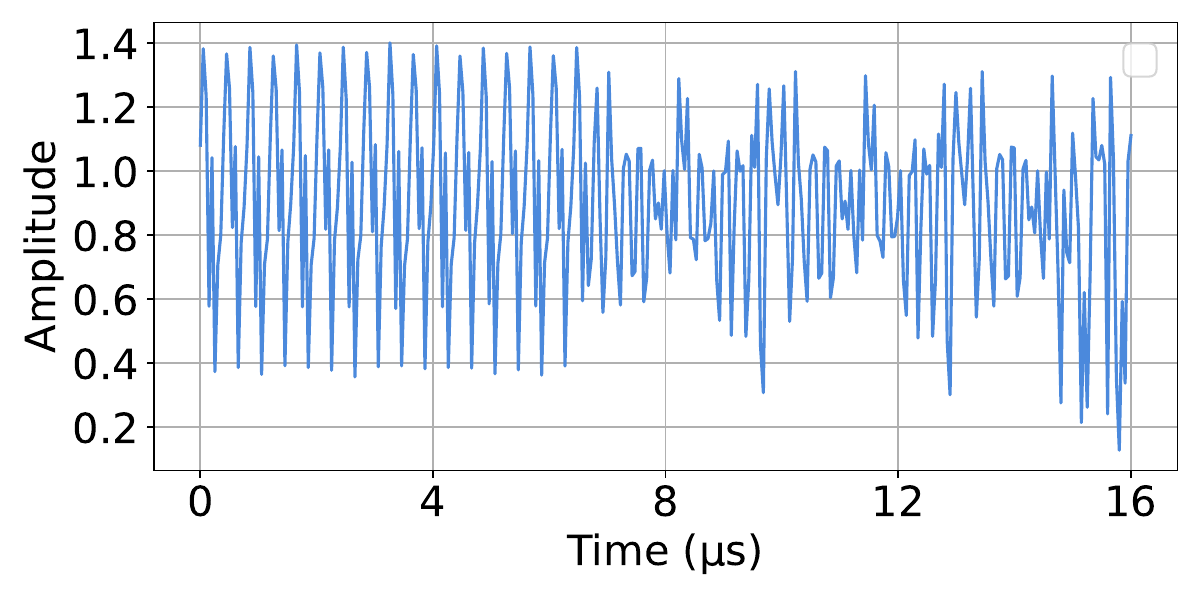}\label{fig:denoised}}
\caption{Exemplification of signal waveforms. (a) Original signal (40~dB). (b) Noised signal (5~dB). (c) Denoised signal from 5~dB. }
\label{fig:waveform}
\end{figure*}

Fig.~\ref{fig:waveform} exemplifies the waveforms of the original signal (40 dB), the noisy signal (5 dB), and the denoised signal from 5 dB. The noisy signal, shown in Fig.~\ref{fig:waveform}(b) shows significant fluctuations and show significant information loss compared to the original signal. The denoised signal, shown in Fig.~\ref{fig:waveform}(c) reveals a substantial improvement, with noise significantly removed and the signal shape more closely resembling that of the original signal. The visual comparison highlights the efficacy of the DM in reconstructing the signal, even under moderate noise conditions.

\begin{figure}[!t]
    \centering
    \includegraphics[width=1\linewidth, height=4cm]{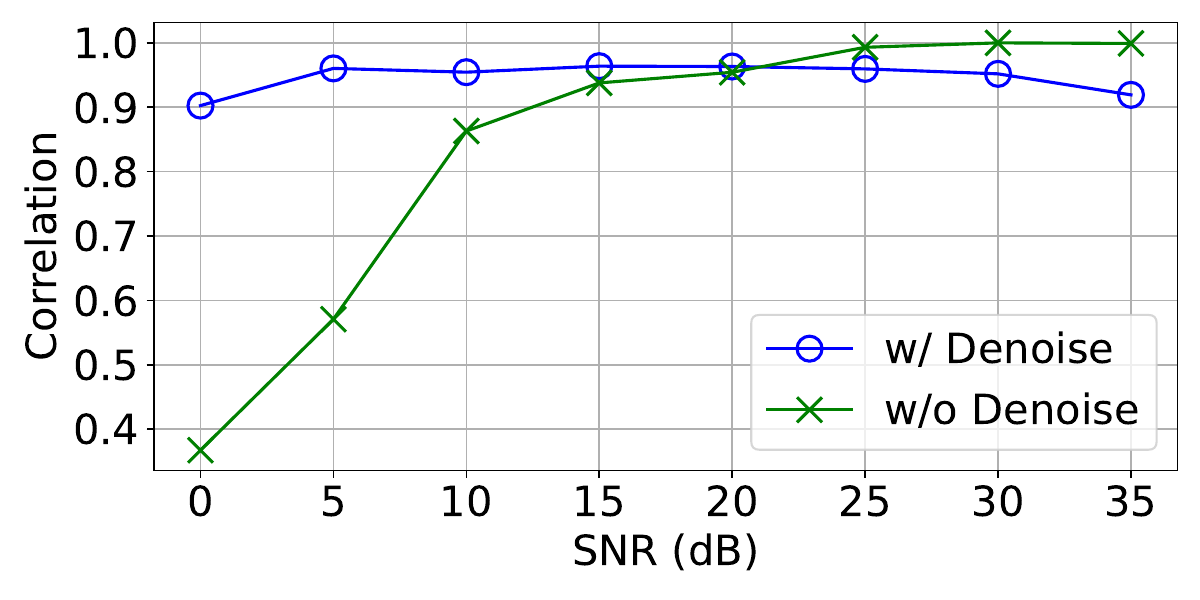}
    \caption{The comparison of the correlation for ``Noised - Original (40dB)" and ``Denoised - Original (40dB)"}
    \label{fig:correlation}
\end{figure}

Fig.~\ref{fig:correlation} shows the correlation between the noise signal and original clean signal, as well as the correlation between the denoised signal and the original signal, at different SNR levels (0~dB to 40~dB). When the SNR is below 20 dB, the DM effectively recovers the original signal, maintaining high correlation between the denoised and original signals even under challenging conditions (0~dB). As SNR rises above 20 dB, noisy signal correlation increases and can match or surpass denoised signal correlation. This is because when the SNR is high, the noise is minimal compared to the signal strength, which results in a marginally lower correlation for the denoised signal compared to the noisy one.

The analysis of both waveform comparisons and correlation results indicates that the DM effectively mitigates noise and restores signal quality, especially under low to moderate SNR conditions. As the SNR increases and the noise becomes less significant, the benefit of denoising diminishes slightly, as reflected in the correlation metrics. Nevertheless, the overall results demonstrate that the DM is highly capable of restoring RF signals for fingerprinting purposes, particularly in challenging noise environments.

\subsection{Device Identification}
In this section, we present the device identification accuracy results of our proposed method. For comparison, we trained a baseline model that incorporates noise augmentation but excludes a noise removal step, meaning no noise removal is performed. The rest of the setup is the same as the proposed method in this paper.

\begin{figure}[!t]
    \centering
    \includegraphics[width=1\linewidth, height=4cm]{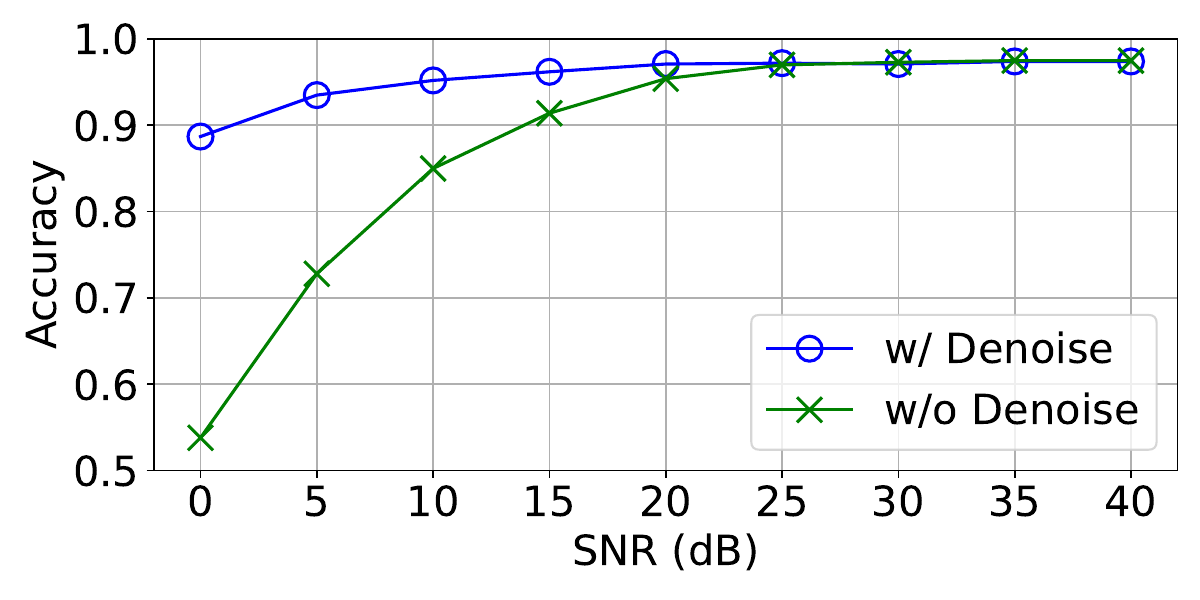}
    \caption{Comparison of the classification performance with and without denoising.}
    \label{fig:acc_compare}
\end{figure}

% This section presents the results of device identification accuracy achieved using the denoised RFF features, as shown in Fig.~\ref{fig:acc_compare}. The comparison includes the fingerprinting performance with and without the denoising step.
The accuracy comparison is illustrated in Fig.~\ref{fig:acc_compare}.
Our proposed approach consistently outperforms the baseline model, when the SNR is below 20~dB. In particular, at an SNR of 0 dB, the proposed model achieves a $34.9\%$ improvement over the baseline. This indicates that the proposed method can significantly enhance the robustness to noise. As the SNR increases beyond 20 dB, the accuracy of both models converges, indicating that the benefit of denoising diminishes when the noise level is low. This convergence is expected, as the impact of noise becomes negligible at higher SNRs, leading to similar performance in both cases. Those observations align with the correlation relationship shown in Fig.~\ref{fig:correlation}.

% However, at low to moderate SNRs, the denoising process substantially enhances the accuracy, highlighting the effectiveness of the DM in mitigating noise and recovering signal features essential for accurate identification. Overall, the results demonstrate that the use of denoised RFF features improves the reliability of device identification under challenging noise conditions, particularly when the SNR is low. The denoising process helps maintain high identification accuracy, making the system more robust to real-world environments where signal quality can vary considerably.

\section{Conclusion}\label{sec:conclude}
In this paper, we present a noise-robust RFFI system by employing a diffusion model to remove noise and restore RFF features. To adapt the diffusion model for noise mitigation within the RFFI system, we propose an SNR mapping method that enables the pretrained noise predictor to estimate and eliminate noise from the signal. To validate the proposed system, we conducted real-world experiments using Wi-Fi as a case study. The experiments demonstrate that the proposed method effectively removes noise and restores RFF. Compared to the simple noise augmentation method, our approach improves the RFFI classification accuracy by up to $34.9\%$.

\section*{Acknowledgement}
The work was supported in part by the UK Engineering and Physical Sciences Research Council (EPSRC) under grant ID EP/Y037197/1 and in part by the UK Royal Society Research Grants RGS$\backslash$R1$\backslash$231435.

\bibliographystyle{IEEEtran}
\bibliography{IEEEabrv,references}
\end{document}